\documentclass{article}

\PassOptionsToPackage{numbers, sort}{natbib}

\usepackage[preprint]{neurips_2025}

\usepackage[utf8]{inputenc} % allow utf-8 input
\usepackage[T1]{fontenc}    % use 8-bit T1 fonts
\usepackage{hyperref}       % hyperlinks
\usepackage{url}            % simple URL typesetting
\usepackage{booktabs}       % professional-quality tables
\usepackage{amsfonts}       % blackboard math symbols
\usepackage{nicefrac}       % compact symbols for 1/2, etc.
\usepackage{microtype}      % microtypography
\usepackage{xcolor}         % colors
\usepackage{enumitem}

\usepackage{amsmath}
\usepackage{amssymb}
\usepackage{mathtools}
\usepackage{amsthm}
\usepackage{multirow}
\usepackage{wrapfig}
\usepackage{graphicx}
\usepackage{subcaption}
\usepackage{wrapfig}
\usepackage{caption} 

\usepackage[capitalize,noabbrev]{cleveref}

\title{Masked Autoencoders that Feel the Heart: \\ Unveiling Simplicity Bias for ECG Analyses}

\author{
  He-Yang Xu$^{1,2}$\thanks{Equal contribution.} \quad
  Hongxiang Gao$^{1,2}$\footnotemark[1] \quad
  Yuwen Li$^{1,2}$ \quad
  Xiu-Shen Wei$^3$\thanks{Corresponding author.} \quad
  Chengyu Liu$^{1,2}$\footnotemark[2] \\
  \small $^1$ State Key Laboratory of Digital Medical Engineering, Southeast University \\
  \small $^2$ School of Instrument Science and Engineering, Southeast University \\
  \small $^3$ School of Computer Science and Engineering, Southeast University \\
  \texttt{\{xuhy, hongxiang\_seu, liyuwen, weixs, chengyu\}@seu.edu.cn}
}

\theoremstyle{plain}

\theoremstyle{definition}

\theoremstyle{remark}

\begin{document}

\maketitle

\begin{abstract}
The diagnostic value of electrocardiogram (ECG) lies in its dynamic characteristics, ranging from rhythm fluctuations to subtle waveform deformations that evolve across time and frequency domains. 
However, supervised ECG models tend to overfit dominant and repetitive patterns, overlooking fine-grained but clinically critical cues—a phenomenon known as Simplicity Bias (SB), where models favor easily learnable signals over subtle but informative ones.
In this work, we first empirically demonstrate the presence of SB in ECG analyses and its negative impact on diagnostic performance, while simultaneously discovering that self-supervised learning (SSL) can alleviate it, providing a promising direction for tackling the bias. 
Following the SSL paradigm, we propose a novel method comprising two key components: 1) Temporal-Frequency aware Filters to capture temporal-frequency features reflecting the dynamic characteristics of ECG signals, and 2) building on this, Multi-Grained Prototype Reconstruction for coarse and fine representation learning across dual domains, further mitigating SB. To advance SSL in ECG analyses, we curate a large-scale multi-site ECG dataset with 1.53 million recordings from over 300 clinical centers. Experiments on three downstream tasks across six ECG datasets demonstrate that our method effectively reduces SB and achieves state-of-the-art performance.

\end{abstract}

\section{Introduction}

\begin{wrapfigure}[17]{r}{0.55\textwidth}
\vspace{-1em}
\centering
\includegraphics[width=0.55\textwidth]{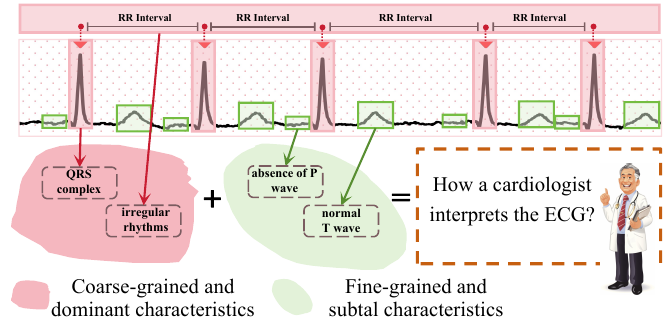}
\caption{\small Illustration of how cardiologists interpret ECG signals by integrating both coarse-grained and fine-grained features. Coarse-grained patterns, such as QRS complexes and RR interval irregularities, provide dominant structural cues. Fine-grained patterns, such as the absence of P waves or subtle T-wave changes, offer essential diagnostic information.}
\vspace{-0.5em}
\label{fig:explain}
\end{wrapfigure}

The electrocardiogram (ECG) remains a cornerstone of non-invasive cardiac assessment, offering a cost-effective and widely accessible method for monitoring  life-threatening conditions such as acute myocardial ischemia \cite{bhatia2018screening, liu2023artificial, lai2023practical, yagi2024routine}. The beauty of ECG lies in its structured patterns encoded across both temporal and frequency domains. As illustrated in Fig.~\ref{fig:explain}, ECG signals exhibit prominent, stable patterns like QRS complexes and rhythm regularities, characterized as low-frequency features representing dominant coarse-grained characteristics. In contrast, rapidly varying and delicate patterns, such as P and T waves—crucial for detecting early-stage or complex cardiac abnormalities—are identified as high-frequency features, encoding subtle fine-grained diagnostic information. Despite the success of powerful supervised models in advancing ECG analyses \cite{wang2024optimizing, gao2023ecg, strodthoff2020deep, alday2020classification, wang2017time}, they often overfit to low-frequency patterns while underrepresenting high-frequency cues critical for clinical interpretation, a phenomenon known as simplicity bias \cite{ shah2020pitfalls, teney2022evading} in supervised learning paradigm.

\begin{figure}[t!]
\centering
\begin{subfigure}[t]{0.62\columnwidth}
    \includegraphics[width=\linewidth]{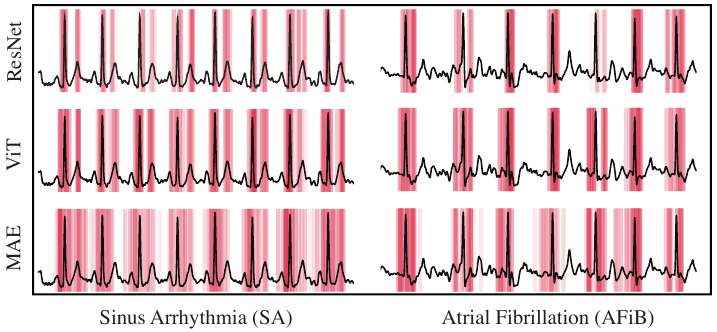}
    \caption{\small{Visualization of supervised vs. self-supervised attention.}}
    \label{fig:cam_a}
\end{subfigure}
\hfill
\begin{subfigure}[t]{0.31\columnwidth}
    \includegraphics[width=1.0\linewidth]{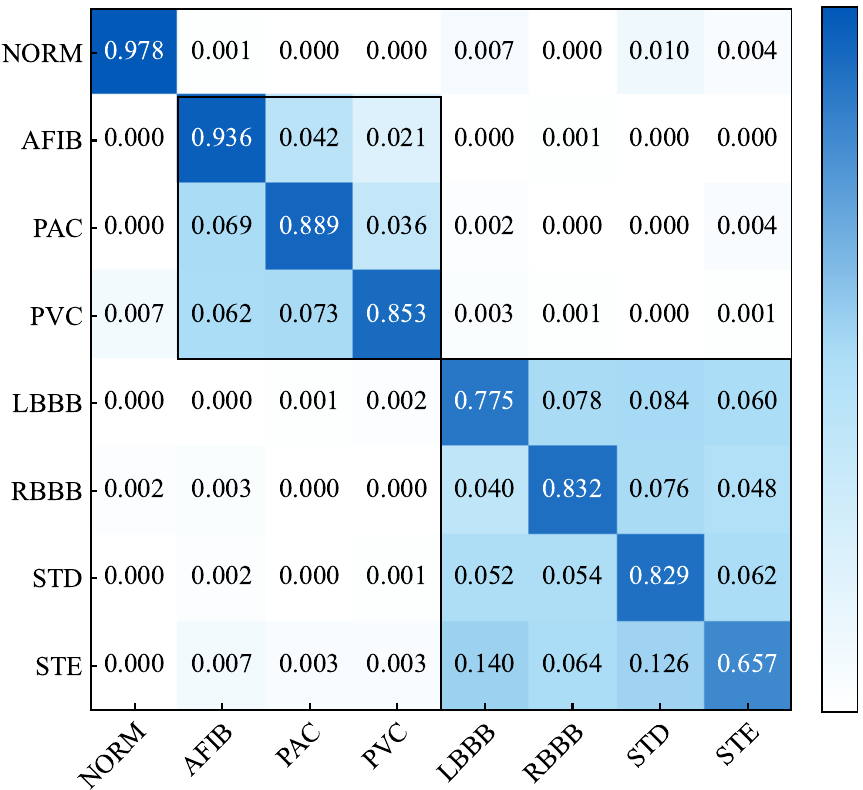}
    \caption{\small{Confusion matrix insights.}}
    \label{fig:cam_b}
\end{subfigure}
\caption{\small{\textbf{Analyses of simplicity bias manifestation in ECG from CPSC2018.} 
(a) SB arises in supervised methods, while self-supervised method better captures high-frequency diagnostic patterns, reducing SB. 
(b) Misclassifications reflect rhythm-level ambiguity in supervised settings.}}
\vskip -0.12in
\end{figure}

Simplicity bias (SB) refers to the tendency of supervised neural networks to prioritize dominant, easily captured features while overlooking subtle but clinically important deviations \cite{shah2020pitfalls, wei2025delving}. Conversely, self-supervised learning (SSL) can mitigate SB by encouraging models to learn more comprehensive representations \cite{wei2025delving, sbssl}. To investigate whether SB manifests in ECG analyses and whether existing ECG SSL methods can alleviate it, we conduct preliminary experiments. As illustrated in Fig.~\ref{fig:cam_a}, supervised models such as ResNet~\cite{wang2017time} and ViT~\cite{vaid2023foundational} tend to focus primarily on dominant QRS complexes and global pseudo-periodic rhythms, while often neglecting subtle high-frequency components such as P and fibrillatory (F) waves. In contrast, the self-supervised model (MAE \cite{zhang2022maefe}) demonstrates partially improved attention to these clinically significant regions, as further discussed in Section \ref{sec:preliminary}. Nonetheless, current ECG SSL methods \cite{vaid2023foundational, zhang2022maefe, hu2023spatiotemporal, wang2023unsupervised, lai2023practical, wang2023adversarial, na2024guiding} are not explicitly designed to address SB. In particular, they lack mechanisms to capture multi-granular representations and model the hierarchical dependencies across temporal-frequency domains, limiting their ability to fully capture the diagnostic richness embedded in ECG signals.

To address such limitation, we propose a novel self-supervised learning method, mu\underline{L}ti-grained t\underline{E}mporal-frequency \underline{A}ware ma\underline{S}ked au\underline{T}oencoders (LEAST). Our proposed method, LEAST, integrates temporal-frequency attention and multi-grained modeling into a masked autoencoder framework, encouraging the self-supervised model to capture challenging features and mitigate SB. The Method consists of two key modules: (1) \underline{T}emporal-\underline{F}requency aware \underline{F}ilters (TFF): We propose multi-head filters within the encoder to capture inherent patterns across both temporal and frequency domains. This design seamlessly integrates temporal and frequency information while capturing domain-specific patterns, enhancing the model's sensitivity to the dynamic nature of ECG signals and effectively mitigating SB. Building on this, (2) \underline{M}ulti-\underline{G}rained \underline{P}rototype \underline{R}econstruction (MGPR): To further extend the model's sensitivity from coarse-grained to fine-grained features, we construct prototypical representations at multiple semantic granularities across dual domains, encouraging the model to preserve information beyond dominant patterns. To advance self-supervised learning research in ECG analyses, we introduce a Chinese multi-site ECG arrhythmia dataset with approximately 380,000 samples collected from 356 clinical centers. Additionally, we integrate multiple publicly available datasets \cite{gow2023mimic, ribeiro2021code} to build a large-scale multi-site ECG dataset, comprising a total of 1,530,000 samples.

Through comprehensive experiments on three downstream tasks—classification, segmentation, and forecasting—we demonstrate that our method is robust across various applications and effectively mitigates the issue of SB. Beyond that, we expect our work to provide a new perspective on ECG analyses grounded in SB, and to contribute toward building a more comprehensive and systematic downstream evaluation framework. Our main contributions are summarized as follows:
\begin{itemize}[leftmargin=2em, itemsep=1pt, topsep=2pt]
\item To the best of our knowledge, we are the first to reveal simplicity bias in ECG analyses and demonstrate its negative impact on representation learning. Based on this insight, we propose LEAST, a self-supervised framework that integrates Temporal-Frequency aware Filters and Multi-Grained Prototype Reconstruction to mitigate the bias.
\item We introduce a large-scale multi-site ECG dataset, which includes our Chinese multi-site ECG arrhythmia dataset comprising 380,000 samples from over 300 clinical centers, annotated across 61 multi-label classes, and an aggregation of existing open-accessed datasets to form a geographically diverse dataset of 1,530,000 samples.
\item LEAST is validated on comprehensive downstream tasks, including classification, segmentation and forecasting, achieving state-of-the-art performance. Visualization results further demonstrate its effectiveness in alleviating simplicity bias.
\end{itemize}

\section{Simplicity Bias Hides the Devil in ECG Analyses}
\label{sec:preliminary}

\subsection{Preliminary Experiments} 

We conduct both qualitative and quantitative experiments to systematically investigate the presence of SB in ECG analyses and its impact on model behavior.

\paragraph{Qualitative analyses} We perform standard supervised learning on CPSC2018 \cite{liu2018open} for ECG classification, using ResNet \cite{wang2017time} as the backbone. Guided by class activation mapping (CAM) \cite{cam}, we visualize the model's attention on an atrial fibrillation (AFiB) signal (lead II used) in Fig.~\ref{fig:cam_a}. The results reveal that the supervised model predominantly focuses on easily recognizable QRS complexes, while largely ignoring subtle regions like P waves and F waves, providing clear evidence of SB in ECG analyses.

\paragraph{Quantitative analyses} We construct supervised binary classification tasks based on eight representative disease categories (except the 1-st atrioventricular block) from the CPSC2018 dataset, selected from the clinical diagnostic perspective (details in Appendix \ref{appendix:sb_setting}). Pairwise classification results are visualized as a confusion matrix in Fig.~\ref{fig:cam_b}. As observed, binary tasks distinguishing between normal and rhythm anomalies, primarily characterized by low-frequency rhythm irregularities, exhibit remarkably low misclassification rates. In contrast, classification within rhythmic or morphological categories demonstrates significantly reduced performance, highlighting the inherent challenges in resolving fine-grained variations. This suggests that current powerful supervised ECG models still struggle to capture subtle high-frequency features critical for distinguishing arrhythmias with overlapping waveform morphologies or rhythm patterns, highlighting the practical implications of simplicity bias in ECG classification.

\subsection{Incorporating Self-Supervised Learning to Mitigate Simplicity Bias}

Based on the previous verification of SB in supervised ECG models, we further evaluate whether SSL can alleviate this issue. We adopt a SSL framework in MAE fashion \cite{zhang2022maefe} and compare it with a supervised baseline, ViT \cite{vaid2023foundational}, under identical settings (see Appendix \ref{appendix:sb_setting}). As shown in Fig. \ref{fig:cam_a}, both models focus on dominant QRS complexes. However, the MAE shows moderately enhanced attention to the presence of P waves, partially mitigating SB. Nonetheless, the improvement remains limited, the ECG SSL model still underattends fine-grained structures critical for diagnosing complex arrhythmias.

These findings suggest that a key challenge remains: enhancing the model's ability to detect fine-grained features while maintaining sensitivity to coarse-grained features, in order to further mitigate SB and advance the reliability of ECG-based health monitoring systems.

\section{Methodology}

% In this section, we propose a novel self-supervised pre-training framework for ECG representation learning, termed as mu\underline{L}ti-grained t\underline{E}mporal-frequency \underline{A}ware ma\underline{S}ked au\underline{T}oencoders (LEAST). Initially, we outline the preliminaries, followed by an overview of LEAST. Subsequently, we delve into the two pivotal modules in subsequent sections.

\begin{figure}[t]
% \vskip 0.2in
\begin{center}
\centerline{\includegraphics[width=\columnwidth]{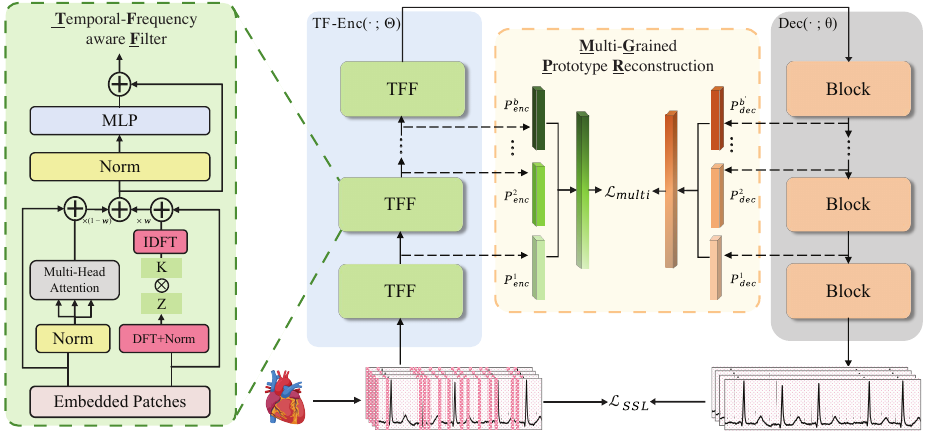}}
\caption{\small{Illustration of the proposed LEAST, which consists of two key components: Temporal-Frequency aware Filter (Section \ref{tfamf}) and Multi-Grained Prototype Reconstruction (Section \ref{mgpr}).}}
\label{fig:piepline}
\end{center}
\vskip -0.3in
\end{figure}

\subsection{Notations}
Given an ECG signal \(X \in \mathbb{R}^{L \times T}\), where \(L\) represents the number of leads and \(T\) denotes the temporal length.  To capture the temporal and morphological patterns inherent in the ECG signal, we first apply a ResNet-like encoder $E(\cdot)$ across all leads simultaneously to capture morphological features and cross-channel dependencies. This results in transformed embeddings along the temporal dimension, represented as \(\boldsymbol{z} = E(X) \in \mathbb{R}^{C \times n}\), where \(C\) is the encoder's output dimensionality, and \(n = \frac{T}{p}\) denotes the number of patches generated from temporal segments of length \(p\). In the self-supervised pre-training phase, we utilize an unlabeled dataset \(\mathcal{D}^{{pre}} = \{x_i^{{pre}} | i = 1, \ldots, N\}\), subsequently fine-tuning on a smaller labeled dataset \(\mathcal{D}^{{tune}} = \{(x_i^{{tune}}, y_i) | i = 1, \ldots, M\}\), with \(M \ll N\), to enhance performance on downstream tasks.

\subsection{LEAST: mu\underline{L}ti-grained t\underline{E}mporal-frequency \underline{A}ware ma\underline{S}ked au\underline{T}oencoders for ECG}

The core idea of our LEAST framework is to alleviate SB in ECG analyses by compelling the model to capture multiple semantic granularities across both temporal and frequency domains. 

% We achieve this through a dual-pronged strategy: (1) explicitly incorporating temporal-frequency representations to capture diagnostic cues that are often overlooked in the time domain, and (2) enforcing multi-grained prototype reconstruction across dual domains to preserve comprehensive features beyond dominant waveform patterns.

As illustrated in Fig.~\ref{fig:piepline}, LEAST consists of two key components: \underline{T}emporal-\underline{F}requency aware \underline{F}ilters (TFF) and \underline{M}ulti-\underline{G}rained \underline{P}rototype \underline{R}econstruction (MGPR). Before these stages, given a standard 12-lead ECG recording \(X \in \mathcal{D}^{pre}\), we use a ResNet-based embedding strategy (\textit{ECG-adapted Patch Embedding}, detailed in Appendix \ref{appendix:patch}) that captures local morphological features and inter-lead correlations, forming structured temporal patches with positional embeddings. Then, we introduce the TFF module, which explicitly models rhythm-related dependencies across both temporal and frequency domains. Building on this, to further reinforce attention to subtle but clinically important characteristics, we propose MGPR. Unlike global patch reconstruction, MGPR aligns intermediate encoder representations at multiple semantic levels, and performs multi-grained reconstruction. After passing through the decoder, the reconstructed signal is aligned with the original ECG input. The overall training objective integrates a global masked reconstruction loss and a multi-grained prototype reconstruction loss across dual domains, encouraging the model to capture fine-grained features beyond dominant coarse-grained patterns, effectively addressing the limitations of simplicity bias. The final training loss is defined as:
\begin{equation}
\label{eqa:mainloss}
\mathcal{L}=\mathcal{L}_{ssl} + \mathcal{L}_{multi}.
\end{equation}

\subsubsection{Temporal-Frequency Aware Filters}
\label{tfamf}

Dynamic is a fundamental attribute of ECG signals. Effectively capturing their waveform patterns and variations demands a dual-pronged analyses in both the temporal and frequency domains. Thus, we propose TFF to capture temporal-frequency features directly within the encoder.

Concretely, to extract frequency-domain features, we first apply the Discrete Fourier Transform (DFT) to the temporal embedding \( \boldsymbol{z} \in \mathbb{R}^{C \times n} \), converting it to the frequency-domain representation:
\begin{equation}
\mathbf{Z}[k] = \sum_{t=0}^{n-1} \boldsymbol{z}[t] \cdot e^{-2\pi i kt/n}, \quad k = 0, \ldots, n-1,
\end{equation}
where \( k \) is the index of the frequency bin, ranging from 0 to \( n-1 \), indicating the specific frequency component of the signal being analyzed. To effectively integrate the tokens in the frequency space, we then compute the query vector \( \mathbf{Q} \in \mathbb{R}^{1 \times D} \) by projecting the real part of \( \mathbf{Z} \) using a learnable matrix \( \mathbf{W}_q \). Then, we introduce multi-head complex filters \( \mathbf{K} \in \mathbb{R}^{H \times D} \), where \( H \) is the number of heads, to compute transformed frequency features via Hadamard product:
\begin{equation}
\tilde{\mathbf{Z}} = \mathbf{Z} \odot (\mathbf{Q} \mathbf{K}) = \mathbf{Z} \odot (\mathbf{Z} \mathbf{W}_q \mathbf{K}).
\end{equation}

After that, we apply Inverse DFT to transform frequency-domain features back into the temporal domain, ensuring compatibility for subsequent fusion with the temporal attention outputs:
\begin{equation}
\boldsymbol{\widehat{z}}[t] = \frac{1}{n} \sum_{k=0}^{n-1} \widetilde{\mathbf{Z}}[k] \cdot e^{2\pi i kt/n},
\end{equation}
where $\boldsymbol{\widehat{z}}$ represents the feature back to temporal domain. In parallel, the original embedding \( \boldsymbol{z} \) is passed through a standard self-attention block to extract temporal-domain features:
\begin{equation}
\boldsymbol{z}_t = \mathrm{SelfAttn}(\boldsymbol{z}).
\end{equation}
We then fuse the dual-view representations using a learnable weight \( w_{{fuse}} \in [0, 1] \):
\begin{equation}
\boldsymbol{z}_{fused} = w_{fuse} \cdot \widehat{\boldsymbol{z}} + (1 - w_{fuse}) \cdot \boldsymbol{z}_t.
\end{equation}
This fused representation serves as the input to subsequent layers, and multiple TFF-enhanced blocks are stacked to construct the encoder $\text{TF-Enc}(\cdot; \Theta)$.

\subsubsection{Multi-Grained Prototype Reconstruction}
\label{mgpr}

To introduce multi-grained information into ECG representation learning and thereby mitigate simplicity bias, we propose MGPR. For each encoder block \(b \in \mathcal{B}_{enc}\) and decoder block \(b' \in \mathcal{B}_{dec}\), we compute semantic prototypes by averaging the output embeddings across all sequences in the batch to obtain robust representations:
\begin{equation}
\boldsymbol{\tilde{P}}_{enc}^{b, i} = \frac{1}{n} \sum_{j=1}^{n} \boldsymbol{z}_{enc}^{b, i, j}, \quad
\boldsymbol{\tilde{P}}_{dec}^{b', i} = \frac{1}{n} \sum_{j=1}^{n} \boldsymbol{z}_{dec}^{b', i, j},
\end{equation}
where \( \boldsymbol{z}_{enc}^{b, i, j} \) and \( \boldsymbol{z}_{dec}^{b', i, j} \) represent the patch-level outputs from the \(b\)-th encoder block and the \(b'\)-th decoder block for the \(i\)-th sample and the \(j\)-th patch, respectively. And $\tilde{P}_{dec}^{b'}$ and $\tilde{P}_{enc}^{b}$ denote the prototype from decoder and encoder one batch. Averaging over all patches within each sample yields a per-sample prototype that summarizes the semantic information at a specific block. Notably, the decoder typically adopts a lightweight design with fewer blocks (\( |\mathcal{B}_{dec}| < |\mathcal{B}_{enc}| \)), resulting in reconstructions aligned only with a subset of encoder prototypes.
To enable comparison across layers with differing output dimensions, we apply learnable linear projections to map encoder and decoder prototypes to a shared representation space:
\begin{equation}
\boldsymbol{P}_{enc}^{b} = \text{Proj}_{enc}(\tilde{P}_{enc}^{b}), \quad
\boldsymbol{P}_{dec}^{b'} = \text{Proj}_{dec}(\tilde{P}_{dec}^{b'}), 
\end{equation}
where \( \text{Proj}_{enc} \) and \( \text{Proj}_{dec} \) are learnable linear mappings that project prototypes from encoder block \(b\) and decoder block \(b'\), respectively, into a shared latent space.

To enforce alignment across semantic levels, we compute the mean squared error (MSE) between the projected prototypes from selected encoder and decoder blocks:
\begin{equation}
\mathcal{L}_{multi} = \frac{1}{|\mathcal{H}|} \sum_{(b, b') \in \mathcal{H}} \left\| \boldsymbol{P}_{enc}^{b} - \boldsymbol{P}_{dec}^{b'} \right\|_2^2,
\end{equation}
where $ \mathcal{H}$ defines the ordered mapping of selected block pairs for multi-level alignment. The overall training objective combines the patch-wise masked reconstruction loss $\mathcal{L}_{ssl}$ with the prototype-based multi-level alignment loss $\mathcal{L}_{multi}$, as defined in Equation~\ref{eqa:mainloss}.

% Existing SSL methods in ECG analyses are typically not designed to capture fine-grained features \cite{zhang2022maefe, wang2023adversarial}, biasing the model toward coarse-grained components. This exacerbates simplicity bias and results in insufficient modeling of subtle patterns. Thus, we propose Multi-Grained Prototype Reconstruction to explicitly enforce attention across semantic levels, encouraging the model to capture features ranging from coarse-grained structures to fine-grained details across dual domains.

% To effectively model features across multiple levels of semantic granularity, we extract prototypes from variable layers throughout the encoder and decoder blocks. 

% where \( \mathcal{H} \subseteq \mathcal{B}_{enc} \times \mathcal{B}_{dec} \) defines the set of block pairs selected for multi-level alignment. 

% This formulation introduces hierarchical alignment signals that regularize learning across semantic levels, complementing standard patch-level objectives. In contrast to conventional approaches that reconstruct only from the final decoder output, MGPR enforces consistency between latent structures at varying depths. This encourages the model to internalize fine-grained diagnostic features often neglected due to SB. The overall training objective combines the patch-wise masked reconstruction loss $\mathcal{L}_{ssl}$ with the prototype-based multi-level alignment loss $\mathcal{L}_{multi}$, as defined in Equation~\ref{eqa:mainloss}.

\section{Large-scale Datasets for ECG Pretraining}
\label{exp:pre_data}
High-quality and diverse ECG datasets are critical for learning robust and generalizable representations in pretraining. To this end, we utilized a diverse set of large-scale, real-world clinical ECG datasets spanning multiple continents, healthcare systems, and patient demographics.

\subsection{Self-Constructed Multi-Center Clinical Dataset}
To support reliable and generalizable ECG representation learning, we construct a comprehensive, high-fidelity clinical ECG dataset comprising over 380,000 12-lead recordings from 356 medical centers across Shandong Province, China, collected over a 10-year span (2012–2022). Each recording is sampled at 500 Hz for 10 seconds and corresponds to a unique patient, totaling more than 300,000 individuals, covering a broad population base. All data and institutional identifiers were de-identified prior to analyses.

\begin{figure}[t]
% \vskip 0.2in
\begin{center}
\centerline{\includegraphics[width=0.85\columnwidth]{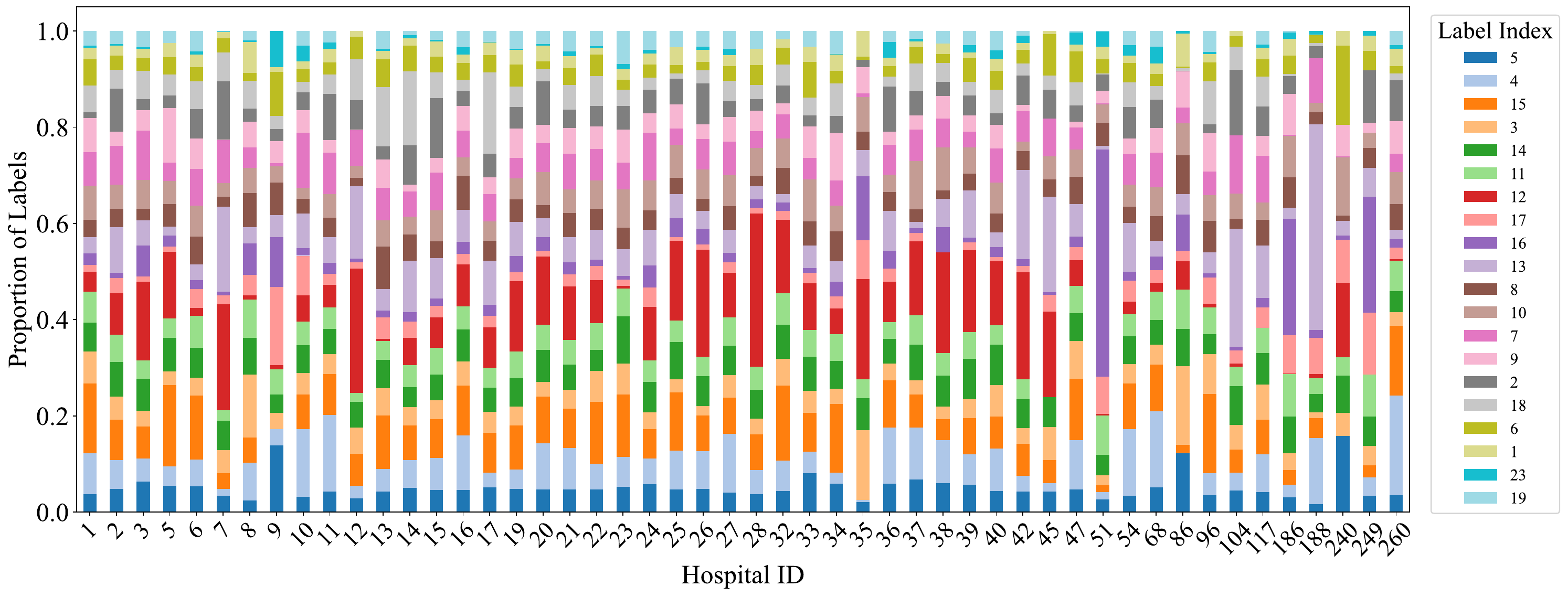}}
\caption{\small{Normalized distribution of top 20 diagnostic labels across top 50 centers.}}
\label{fig:class_dist}
\end{center}
\vskip -0.3in
\end{figure}

% \begin{wrapfigure}[11]{r}{0.63\textwidth}
%     \vspace{0em}
%     \centering
%     \includegraphics[width=0.63\textwidth]{figs/Age and Sex Distribution per Hospital.pdf}
%     \caption{\small Age and Sex Distribution per Hospital (Male/Female).}
%     \label{fig:geo_dist}
%     \vspace{-1em}
% \end{wrapfigure}

The dataset reflects substantial real-world heterogeneity, encompassing rural clinics, county hospitals, and tertiary centers, and introduces variation in acquisition devices, demographics, and clinical protocols. All ECGs were acquired through a remote monitoring system and centrally annotated by at least two board-certified cardiologists using standardized criteria, with ambiguous cases resolved through expert consensus. This multi-reader protocol ensures trustworthy, fine-grained labels, supporting high-quality supervision for representation learning.

Diagnostic labels span 61 clinically meaningful categories, including rhythm disorders, conduction abnormalities, ischemic events, hypertrophy, and axis deviations. Importantly, the dataset retains real-world distributional bias, such as class imbalance across different center ( Fig. \ref{fig:class_dist}) and demographic skew, providing a robust benchmark for evaluating generalization under realistic clinical variability.

\subsection{Complementary Public ECG Datasets}
To enhance the generalizability of pretraining beyond our own clinical setting, we integrate two large-scale and complementary open-access ECG datasets.

\textbf{MIMIC-IV-ECG} \cite{gow2023mimic} includes approximately 800,000 12-lead recordings from 160,000 patients collected between 2008–2019 at the Beth Israel Deaconess Medical Center in Boston, USA. As a representative of high-acuity inpatient and ICU care in a developed country, it offers valuable insight into ECG patterns under intensive monitoring.

\textbf{CODE-15\%} \cite{ribeiro2021code} contains 345,779 12-lead ECGs from 233,770 patients across hundreds of municipalities in Brazil. This dataset reflects low-resource outpatient and primary care settings, contributing substantial geographic and epidemiological diversity.

The integration of these datasets with our self-constructed multi-center corpus yields a cross-continental, cross-setting, and cross-population pretraining dataset, encompassing ICU to rural clinics, and patients from diverse socioeconomic and ethnic backgrounds. This rich coverage forms a comprehensive and high-fidelity foundation for learning generalizable ECG representations across varied clinical environments.
% \begin{table}[t]
% \centering
% \caption{Summary of ECG datasets used for pretraining and evaluation.}
% \label{tab:dataset_summary}
% \begin{tabular}{lcccccc}
% \toprule
% \textbf{Dataset} & \textbf{Source} & \textbf{\#Records} & \textbf{Sampling Rate} & \textbf{Setting} & \textbf{Annotation} & \textbf{Use} \\
% \midrule
% Ours & China (356 centers) & 380,000 & 500 Hz & Rural to Tertiary & Expert-reviewed (2x) & Pretrain \\
% MIMIC-IV-ECG & USA (BIDMC ICU) & 800,000 & 500 Hz & ICU / Inpatient & Automated + Notes & Pretrain \\
% CODE-15\% & Brazil (TNMG) & 345,779 & 400 Hz & Outpatient / Primary & Clinician-reviewed & Pretrain \\
% \bottomrule
% \end{tabular}
% \end{table}

\begin{table}[h]
\vspace{-0.8em}
\centering
\footnotesize
\renewcommand{\arraystretch}{1.1}
\setlength{\tabcolsep}{4.85pt}
\caption{Comparative summary of ECG datasets used for pretraining. }
\label{tab:dataset_comparison}
\begin{tabular}{lcccccc}
\toprule
\textbf{Dataset} & \textbf{Source} & \textbf{\#Patients} & \textbf{\#Records}  & \textbf{Setting} & \textbf{Span (Years)} \\
\midrule
Ours  & China (356 centers) & 380,000 & 380,000 & Rural to Tertiary & 2012--2022 \\
MIMIC-IV-ECG       & USA (BIDMC ICU) & 160,000 & 800,000  & ICU / Inpatient & 2008--2019 \\
CODE-15\%          & Brazil (TNMG) & 233,770 & 345,779  & Outpatient / Primary & 2010--2016 \\
\bottomrule
\end{tabular}
\vskip -0.1in
\end{table}

\section{Experiments}
\subsection{Datasets and Empirical Settings}
\label{sec:es}
To advance research on self-supervised pretraining in ECG analyses, we constructed large-scale multi-site ECG pretraining dataset. It includes 380,000 proprietary samples collected from over 356 hospitals, fully annotated with 61 diagnostic categories, as well as curated public datasets \cite{gow2023mimic, ribeiro2021code}, resulting in a combined corpus of 1,530,000 samples. For fair comparison, all self-supervised baselines \cite{zhang2022maefe, Yuqietal-2023-PatchTST, mae, chen2021empirical, wang2023unsupervised, na2024guiding} are re-pretrained on this dataset in our experiments. 

For downstream adaptation, we adopt two paradigms: linear probing and full fine-tuning. We conduct experiments across three downstream tasks—classification, segmentation (detecting characteristic waveform components, \textit{i.e.} R-peaks locating task) and forecasting (predicting survival time from ECG signals, a form of survival analyses)—on six widely used ECG datasets. More details about the task definition, downstream datasets and metrics could be seen in Appendix \ref{appendix:expsetup1}. 
For a fair comparison, following \cite{vaid2023foundational, zhang2022maefe,  na2024guiding}, all experiments are conducted using the ViT-Base encoder \cite{vaid2023foundational}, pretrained on our curated ECG dataset. The empirical settings for training are detailed in the Appendix \ref{esetting}. 

\subsection{Main Results}
We evaluate our method across three downstream tasks: classification, segmentation, and forecasting. Detailed task descriptions and evaluation metrics are provided in Appendix \ref{appendix:expsetup1}.

\paragraph{Comparisons on Classification} We conduct classification experiments on PTB-XL \cite{strodthoff2020deep}, Ningbo \cite{zheng2020optimal}, Chapman \cite{zheng202012}, and CPSC2018 \cite{liu2018open}, covering a wide range of clinical conditions, geographic regions, and patient demographics. We report the results in Table \ref{tab:ecg_classification}. Results show that our LEAST method consistently achieves significant accuracy improvements over state-of-the-art methods.
% It is worth noting that, as mentioned above, CPSC2018 contains AFiB, a condition particularly vulnerable to simplicity bias in ECG analyses. Consequently, our method achieves even greater improvements over state-of-the-art approaches on this dataset.

\begin{table}[t!]
\footnotesize
\centering
\renewcommand\arraystretch{1.1}
\setlength{\tabcolsep}{4.2pt} 
\caption{\small Classification performance across four datasets (all metrics in \%). Best results are in \textbf{bold}.}
\label{tab:ecg_classification}
\begin{tabular}{
c|
@{\hspace{6pt}}c@{\hspace{2pt}}c@{\hspace{2pt}}c|
c@{\hspace{2pt}}c@{\hspace{2pt}}c|
c@{\hspace{2pt}}c@{\hspace{2pt}}c|
c@{\hspace{2pt}}c@{\hspace{2pt}}c
% @{\hspace{6pt}}c@{\hspace{2pt}}c@{\hspace{2pt}}c@{\hspace{6pt}}|
% @{\hspace{6pt}}c@{\hspace{2pt}}c@{\hspace{2pt}}c@{\hspace{6pt}}|
% @{\hspace{6pt}}c@{\hspace{2pt}}c@{\hspace{2pt}}c
}
\toprule
\multirow{2}{*}{\textsc{Method}} 
& \multicolumn{3}{c|}{\emph{PTB-XL}} 
& \multicolumn{3}{c|}{\emph{Ningbo}} 
& \multicolumn{3}{c|}{\emph{Chapman}} 
& \multicolumn{3}{c}{\emph{CPSC2018}} 
\\
& Acc & AUROC & F1 
& Acc & AUROC & F1 
& Acc & AUROC & F1 
& Acc & AUROC & F1 
\\
\midrule
Supervised \cite{vaid2023foundational} & 88.47 & 92.40 & 72.39 & 86.58 & 96.40 & 87.67 & 83.60 & 95.73 & 85.66 & 95.32 & 96.10 & 77.00 \\
\midrule
\multicolumn{13}{c}{\textit{Linear Probing}} \\
\hline
MAE \cite{mae}            & 87.64 & 91.37 & 70.48 & 73.16 & 89.25 & 76.54 & 73.31 & 89.36 & 75.35 & 94.81 & 95.68 & 76.11 \\
MOCO v3 \cite{chen2021empirical}        & 86.76 & 90.92 & 67.17 & 73.40 & 90.12 & 76.11 & 74.41 & 86.13 & 74.93 & 92.37 & 93.70 & 70.42 \\
MTAE \cite{zhang2022maefe}           & 86.79 & 89.46 & 66.41 & 72.89 & 88.49 & 76.63 & 75.33 & 89.44 & 76.24 & 95.15 & 95.46 & 73.00 \\
MLAE \cite{zhang2022maefe}         & 86.68 & 88.87 & 63.88 & 71.61 & 88.35 & 74.35 & 69.69 & 88.26 & 71.12 & 94.16 & 94.90 & 65.67 \\
PatchTST \cite{Yuqietal-2023-PatchTST}       & 85.37 & 86.27 & 64.54 & 72.20 & 87.18 & 73.41 & 67.92 & 85.03 & 68.87 & 94.33 & 95.00 & 70.38 \\
ST-MEM \cite{na2024guiding}        & 87.33 & 90.30 & 66.39 & 73.25 & 89.22 & 76.03 & 73.12 & 89.91 & 75.44 & 95.24 & 95.42 & 75.17 \\
\hline
\textbf{LEAST} & \textbf{88.25} & \textbf{92.07} & \textbf{71.65} & \textbf{73.71} & \textbf{90.41} & \textbf{76.70} & \textbf{76.19} & \textbf{90.11} & \textbf{76.41} & \textbf{95.64} & \textbf{96.12} & \textbf{76.49} \\
\midrule
\multicolumn{13}{c}{\textit{Fine-tuning}} \\
\midrule
MAE \cite{mae} & 88.81 & 93.00 & 73.36 & 86.01 & 96.64 & 86.70 & 88.20 & 97.33 & 89.61 & 96.60 & 97.23 & 84.50 \\
MOCO v3 \cite{chen2021empirical}  & 87.31 & 92.27 & 71.45 & 86.55 & 96.42 & 84.39 & 85.25 & 96.61 & 87.12 & 95.28 & 95.29 & 79.24\\
MTAE \cite{zhang2022maefe}    & 87.39 & 90.33 & 69.36 & 86.31 & 95.52 & 87.45 & 85.94 & 97.70 & 85.06 & 95.37 & 95.73 & 83.04 \\
MLAE \cite{zhang2022maefe}    & 88.12 & 90.28 & 64.94 & 85.94 & 96.14 & 86.78 & 86.94 & 97.48 & 84.43 & 95.02 & 96.60 & 79.78 \\
PatchTST \cite{Yuqietal-2023-PatchTST} & 86.83 & 89.30 & 66.36 & 82.75 & 95.68 & 87.12 & 87.19 & 97.39 & 83.85 & 95.00 & 96.09 & 83.11 \\
ST-MEM \cite{na2024guiding}  & 88.56 & 92.69 & 72.88 & 86.73 & 96.47 & 85.17 & 87.45 & 97.71 & 89.70 & 95.02 & 95.72 & 84.51 \\
\hline
\textbf{LEAST} & \textbf{89.60} & \textbf{93.60} & \textbf{74.70} & \textbf{86.88} & \textbf{97.04} & \textbf{87.71} & \textbf{88.09} & \textbf{98.12} & \textbf{89.32} & \textbf{97.55} & \textbf{98.03} & \textbf{85.19} \\
\bottomrule
\end{tabular}
\vskip -0.1in

\end{table}

\begin{table}[t!]
\footnotesize
\centering
\renewcommand\arraystretch{1.1}
\setlength{\tabcolsep}{4.85pt}
\caption{\small Segmentation and forecasting performance (all metrics in \%). Best results are in \textbf{bold}.}
\label{tab:segmentation_forecasting}
\begin{tabular}{c|ccc|ccc|cc|cc}
\toprule
\multirow{3}{*}{\textsc{Method}} 
& \multicolumn{6}{c|}{\emph{Segmentation (CPSC-QRS)}} 
& \multicolumn{4}{c}{\emph{Forecasting (SAMI-TROP)}} \\
\cline{2-11}
& \multicolumn{3}{c|}{\emph{Linear Probing}} 
& \multicolumn{3}{c|}{\emph{Fine-tuning}} 
& \multicolumn{2}{c|}{\emph{Linear Probing}}
& \multicolumn{2}{c}{\emph{Fine-tuning}}  \\
& Se & PPV & F1 & Se & PPV & F1 & C-index$\uparrow$ & Brier$\downarrow$ & C-index$\uparrow$ & Brier$\downarrow$ \\
\midrule
Supervised \cite{vaid2023foundational} & - & - & - & 99.26 & 99.66 & 99.46  & - & - & 0.7443 & 0.1481\\
\midrule
% MAE-tiny & 99.30 & 99.62 & 99.46 & 0.7136 \\
MAE \cite{mae} & 99.15 & 99.56 & 99.36 & 99.30 & 99.62 & 99.46  & 0.7398 & 0.1498 & 0.7673 & 0.1089\\
MTAE \cite{zhang2022maefe}    & 97.62 & 99.19 & 98.40 & 99.17 & 99.35 & 99.26  & 0.7426 & 0.1100 & 0.7581 & 0.1394\\
% MTAE     & 97.62 & 99.19 & 98.40 & 99.17 & 99.35 & 99.26  & - \\
MLAE \cite{zhang2022maefe}     & 94.45 & 98.54 & 96.45 & 98.01 & 98.92 & 98.46  & 0.7612 & 0.1155 & 0.7912 & 0.1641\\
PatchTST \cite{Yuqietal-2023-PatchTST} & 89.91 & 98.13 & 98.58 & 99.12 & 99.40 & 99.28  & 0.7347 & 0.1089 & 0.7560 & 0.1487 \\
ST-MEM \cite{na2024guiding}   & 89.93 & 98.02 & 97.00 & 99.19 & 99.57 & 99.38  & 0.7527 & 0.1230 & 0.7623 & 0.0915 \\
\hline
\textbf{LEAST} & \textbf{99.40} & \textbf{99.69} & \textbf{99.48}& \textbf{99.68} & \textbf{99.73} & \textbf{99.71} & \textbf{0.7763} & \textbf{0.0886} & \textbf{0.8066} & \textbf{0.0815}\\
\bottomrule
\end{tabular}
\end{table}
\vspace{-0.5em}

\paragraph{Comparisons on Segmentation and Forcasting} Unlike existing SSL methods that primarily focus on classification, we provide a more comprehensive downstream evaluation by additionally conducting segmentation and forecasting tasks on CPSC-QRS \cite{gao2023ecg} and SAMI-TROP \cite{ribeiro2021sami}, respectively. As shown in Table~\ref{tab:segmentation_forecasting}, our method achieves substantial improvements across both tasks. Notably, despite the near-saturation performance of baseline methods on the segmentation task, our approach yields consistent and measurable improvements. These improvements stem from our reconstruction of multi-granularity representations in the frequency domain, guiding the model to capture comprehensive diagnostic features and enabling more generalizable feature learning across diverse downstream tasks.

\subsection{Ablation Studies}
\paragraph{Effectiveness of TFF and MGPR Components} To evaluate the effectiveness of the two key components in our framework, we conduct ablation studies on classification tasks with full fine-tuning. As shown in Table~\ref{tab:ablation_classification}, different configurations are constructed by incrementally adding TFF and MGPR to the MAE baseline. It is worth noting that the baseline already incorporates our task-specific patchification strategy. Beyond this, we observe consistent performance improvements from both TFF and MGPR, with the full model achieving the highest scores. Individually adding either module yields results that are competitive with state-of-the-arts, with TFF contributing a more pronounced improvement, due to its frequency-aware design. These quantitative gains suggest that our components effectively mitigate the impact of SB in ECG analyses.

\vspace{-0.65em} 
\begin{table}[h]
\caption{\small Ablation study of TFF and MGPR on classification (all metrics in \%). Best results are in \textbf{bold}.}
\label{tab:ablation_classification}
\centering
\footnotesize
\renewcommand{\arraystretch}{1.1}
\setlength{\tabcolsep}{5pt}
\begin{tabular}{c|ccc|ccc}
\toprule
\multirow{2}{*}{\textbf{Method}} & \multicolumn{3}{c|}{\textit{PTB-XL}} & \multicolumn{3}{c}{\textit{CPSC2018}} \\
 & Accuracy & AUROC & F1 & Accuracy & AUROC & F1 \\
\midrule
MAE             & 88.81 & 93.00 & 73.36 & 96.60 & 97.23 & 84.50 \\
+ TFF           & 89.35 & 93.46 & 74.38 & 97.31 & 97.66 & 84.83 \\
+ MGPR          & 89.10 & 93.31 & 74.22 & 97.06 & 97.51 & 84.65 \\
\hline
\textbf{Our LEAST} & \textbf{89.60} & \textbf{93.60} & \textbf{74.70} & \textbf{97.55} & \textbf{98.03} & \textbf{85.19} \\
\bottomrule
\end{tabular}
% \vskip -0.1in
\end{table}

\paragraph{Visual Analyses of Simplicity Bias Mitigation}

\begin{figure}[t]
\centering
\resizebox{1\textwidth}{!}{ 
    \begin{minipage}[t]{0.43\textwidth}
        \centering
        \includegraphics[width=\linewidth]{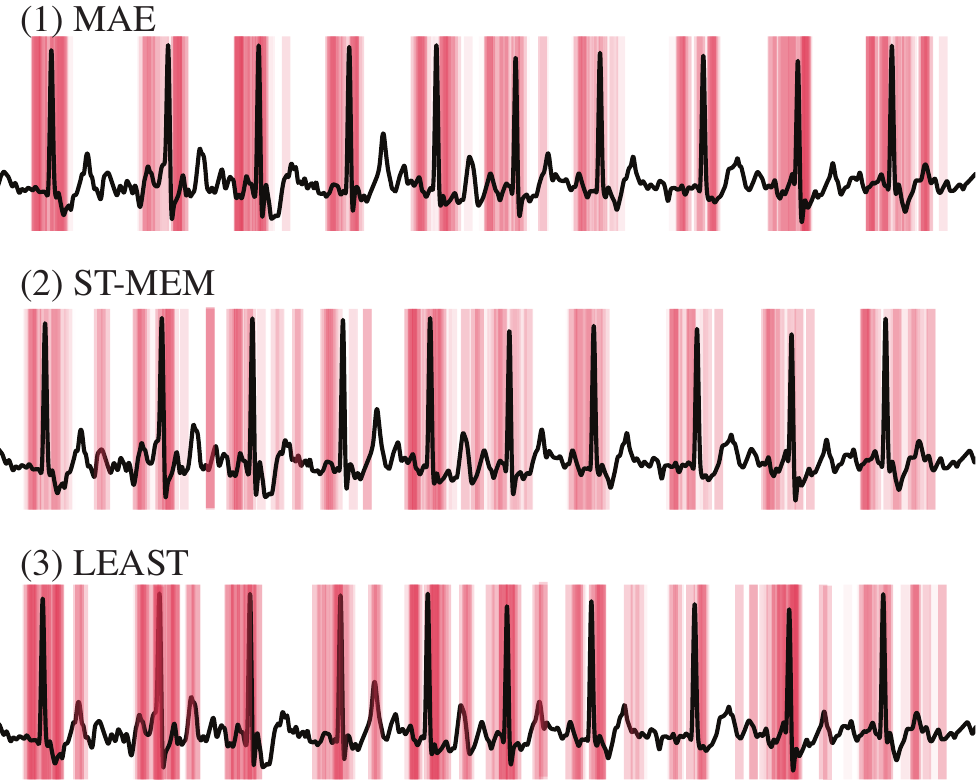}
        \caption{\small Visual comparison of model attention on CPSC2018 samples.}
        \label{fig:cam_visualization}
    \end{minipage}
    \hspace{0.5em} %
    \begin{minipage}[t]{0.5\textwidth}
        \centering
        \includegraphics[width=\linewidth]{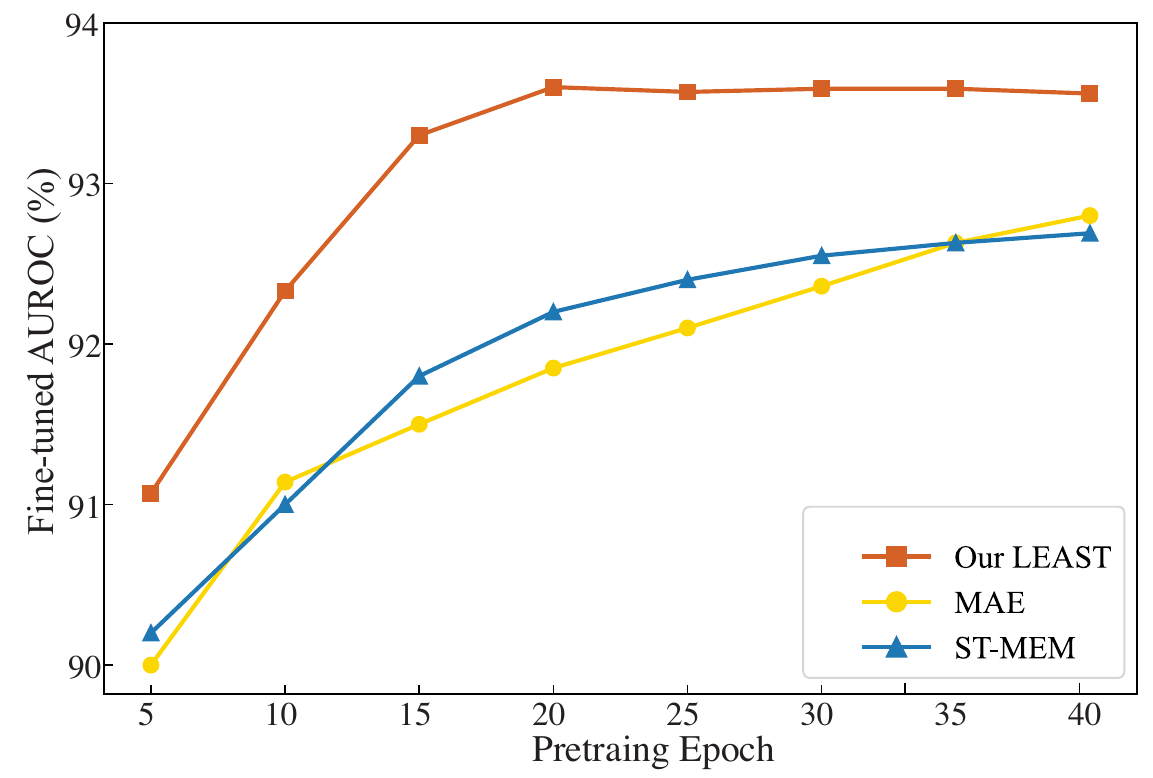}
        \caption{\small AUROC performance with varying pretraining epochs.}
        \label{fig:efficiency}
    \end{minipage}
}
% \vskip -0.2in
\end{figure}

Having established the presence of simplicity bias in ECG analyses and shown that standard SSL methods provide only partial mitigation, we further conduct a visual comparison between our proposed LEAST and existing ECG SSL approaches. Specifically, we employ CAM to examine how different models attend to critical waveform regions. Using representative samples from CPSC2018 (with lead II as an example), we highlight challenging cases such as AFiB. As illustrated in Fig.~\ref{fig:cam_visualization}, we compare MAE~\cite{zhang2022maefe}, ST-MEM~\cite{na2024guiding}, and our LEAST. Notably, LEAST demonstrates markedly improved attention to the structured P–QRS–T waveform, effectively identifying the presence of P waves and subtle deviations of F waves from the counterparts. This visual evidence further validates LEAST’s capability to alleviate simplicity bias and enhance fine-grained feature extraction in complex ECG signals.
% \begin{wrapfigure}[18]{r}{0.5\textwidth}
% \vspace{-1em}
% \centering
% \includegraphics[width=0.5\textwidth]{fig4.pdf}
% \caption{\small Visual comparison of model attention on CPSC2018 samples.}
% \label{fig:cam_visualization}
% \end{wrapfigure}

\paragraph{Learning Efficiency Analyses}

To validate whether our design, which targets temporal-frequency and multi-granularity structures, can more effectively learn robust ECG representations, we conduct 

% \begin{wrapfigure}[14]{l}{0.48\textwidth}
% \vspace{-1em}
% \centering
% \includegraphics[width=0.46\textwidth]{efficiency.pdf}
% \vspace{-0.3em}
% \caption{\small {AUROC performance with varying pretraining epochs.}}
% \label{fig:efficiency}
% \end{wrapfigure}

experiments to evaluate the learning efficiency during the self-supervised pretraining. We fine-tune the models using weights obtained from different pretraining stages on the PTB-XL dataset and report the downstream AUROC performance. We compare the learning efficiency of MAE and ST-MEM with our LEAST framework. As shown in Fig.~\ref{fig:efficiency}, our LEAST achieves peak downstream classification performance after approximately 20 epochs of pretraining, whereas ST-MEM and MAE require around 35 and 40 epochs, respectively. This shows a notable improvement in learning efficiency, highlighting LEAST's enhanced capability to acquire stable ECG representations.

\vspace{-0.9em}
\paragraph{Effectiveness of Prototype-Based Alignment in MGPR}
\label{appendix:ablation2}

\begin{wraptable}[9]{r}{7.6cm}
\footnotesize
\centering
\vspace{-1.3 em}
\renewcommand{\arraystretch}{1.2}  % <-- 这里控制行距
\caption{\small Comparisons between prototype-based and batch-level alignment (all metrics in \%). Best results are in \textbf{bold}.}
\begin{tabular}{c|ccc}
\toprule
\multicolumn{1}{c|}{\multirow{2}{*}{\textsc{Settings}}} & \multicolumn{3}{c}{\emph{CPSC2018}} \\
\multicolumn{1}{c|}{} & Accuracy & AUROC & F1 \\
\hline
Ours (batch-level) & 97.10 & 97.61 & 84.78 \\
Ours (prototype-based) & \textbf{97.55} & \textbf{98.03} & \textbf{85.19} \\
\bottomrule
\end{tabular}
\label{tab:prototype_ablation_final}
% \vspace{-1em}
\end{wraptable}
Prototypical features are known for enhancing feature robustness \cite{snell2017prototypical, duan2024long}. In MGPR, feature alignment between encoder and decoder blocks is achieved through prototype representations, providing more stable features compared to batch-level matching. As shown in Table \ref{tab:prototype_ablation_final}, the results demonstrate the superiority of prototype-based matching over batch-level methods. Importantly, the alignment follows a shallow-to-deep strategy based on relative block depth, with selected encoder blocks distributed across varying depths to enhance diversity and avoid concentration. To explore different encoder block configurations, we conduct experiments using our model with 12 encoder blocks and 8 decoder blocks, as shown in Table \ref{tab:prototype_ablation_final2}. The results indicate that as long as the shallow-to-deep order is maintained and the selections are well-distributed, variations in block choices have minimal impact on performance. This confirms that MGPR's alignment strategy is robust, flexible, and effective without rigid block assignments.

\vspace{-1.1em}
\begin{table}[h]
\caption{\small AUROC results (\%) with randomly selected, well-distributed encoder blocks. Best results are in \textbf{bold}.}
\label{tab:prototype_ablation_final2}
\centering
\footnotesize
\renewcommand{\arraystretch}{1.1}
\setlength{\tabcolsep}{5pt}
\begin{tabular}{c|c|c}
\toprule
\multirow{1}{*}{\textbf{Encoder Block Selection}} & {\textit{PTB-XL}} & {\textit{CPSC2018}} \\
\midrule

\textit{1 $\rightarrow$ 2 $\rightarrow$ 4 $\rightarrow$ 5 $\rightarrow$ 6 $\rightarrow$ 8 $\rightarrow$ 9 $\rightarrow$ 10}          & 93.59  & 98.00  \\
\textit{1 $\rightarrow$ 3 $\rightarrow$ 4 $\rightarrow$ 7 $\rightarrow$ 8 $\rightarrow$ 9 $\rightarrow$ 10 $\rightarrow$ 11}          & 93.58  & \textbf{98.05} \\
\textit{0 $\rightarrow$ 2 $\rightarrow$ 3 $\rightarrow$ 5 $\rightarrow$ 7 $\rightarrow$ 8 $\rightarrow$ 10 $\rightarrow$ 11}  & \textbf{93.60} & 98.03 \\

\bottomrule
\end{tabular}
\vskip -0.1in
\end{table}

\section{Conclusion}
\label{sec:conc}
In this work, we investigated simplicity bias (SB) in ECG analyses, revealing its presence and impact in ECG supervised models, while also discovering that self-supervised learning can help mitigate this bias. Therefore, to further mitigate SB, we adopted the SSL paradigm and proposed our LEAST, which consists of two key components. Temporal-Frequency aware Filters, which integrated time and spectral modeling through learnable multi-head filters, enabling inherent temporal-frequency representation and prioritizing diagnostically critical features across temporal morphologies and frequency oscillations. Building on this, Multi-Grained Prototype Reconstruction enforced multi-level semantic alignment, preserving representations from coarse-grained patterns to fine-grained features across dual domains. To further boost the SSL in ECG analyses, we presented a large-scale multi-site ECG dataset, comprising 1.53 million samples from over 300 clinical centers. Extensive experiments across three tasks and six datasets demonstrated that LEAST effectively mitigated SB and consistently outperformed state-of-the-arts. We hope this work can offer new insights into ECG representation learning through the lens of SB, contributing to the development of a more systematic evaluation framework for self-supervised ECG analyses, and ultimately advances the broader goal of intelligent healthcare. In the future, we plan to explore the impact of SB on long-tail ECG data to enhance detection of rare cardiac events.

% \begin{table}
%   \caption{Sample table title}
%   \label{sample-table}
%   \centering
%   \begin{tabular}{lll}
%     \toprule
%     \multicolumn{2}{c}{Part}                   \\
%     \cmidrule(r){1-2}
%     Name     & Description     & Size ($\mu$m) \\
%     \midrule
%     Dendrite & Input terminal  & $\sim$100     \\
%     Axon     & Output terminal & $\sim$10      \\
%     Soma     & Cell body       & up to $10^6$  \\
%     \bottomrule
%   \end{tabular}
% \end{table}

% 
\bibliographystyle{unsrt}

\bibliography{ref}
% \bibliographystyle{plain}

% \section*{References}

% References follow the acknowledgments in the camera-ready paper. Use unnumbered first-level heading for
% the references. Any choice of citation style is acceptable as long as you are
% consistent. It is permissible to reduce the font size to \verb+small+ (9 point)
% when listing the references.
% Note that the Reference section does not count towards the page limit.
\medskip

% {
% \small

% [1] Alexander, J.A.\ \& Mozer, M.C.\ (1995) Template-based algorithms for
% connectionist rule extraction. In G.\ Tesauro, D.S.\ Touretzky and T.K.\ Leen
% (eds.), {\it Advances in Neural Information Processing Systems 7},
% pp.\ 609--616. Cambridge, MA: MIT Press.

% [2] Bower, J.M.\ \& Beeman, D.\ (1995) {\it The Book of GENESIS: Exploring
%   Realistic Neural Models with the GEneral NEural SImulation System.}  New York:
% TELOS/Springer--Verlag.

% [3] Hasselmo, M.E., Schnell, E.\ \& Barkai, E.\ (1995) Dynamics of learning and
% recall at excitatory recurrent synapses and cholinergic modulation in rat
% hippocampal region CA3. {\it Journal of Neuroscience} {\bf 15}(7):5249-5262.
% }

%%%%%%%%%%%%%%%%%%%%%%%%%%%%%%%%%%%%%%%%%%%%%%%%%%%%%%%%%%%%
\newpage
\appendix

\section{Simplicity Bias Lurking in ECG Analyses}
% \label{appendix:sb}

\subsection{Preliminary Experimental Dataset and Settings}
\label{appendix:sb_setting}

\paragraph{Dataset} 
% \label{appendix:sb_data}
To investigate SB in ECG analyses, we employ the well-annotated CPSC2018 \cite{liu2018open}, which consists of 9 dignostic categories, grouped into three main classes: normal, morphological abnormalities, and rhythm anomalies. Among these, AFiB represents a critical condition characterized by absolute rhythm irregularities, which can often be distinguished from morphological arrhthmias such as left/right bundle branch block (LBBB/RBBB). However, differentiating AFiB from other rhythmic arrhythmias, which also exhibit irregularities, requires capturing high-frequency but subtle waveform variations (absence of P wave and occurance of fibrillatory (F) waves). Therefore, CPSC2018 serves as a suitable dataset for verifying whether SB arises in ECG analyses.

\paragraph{Settings} 
To demonstrate the effectiveness of self-supervised learning in mitigating the simplicity bias (SB) phenomenon, we provide a supervised baseline with an identical setup. Specifically, we adopt ViT-Base \cite{vaid2023foundational} as the backbone and perform both pretraining and finetuning under the same data configurations as the self-supervised setting. Additionally, we select eight categories from the CPSC2018 dataset and construct multiple binary classification tasks by pairing them in all possible combinations. The misclassification rates for each pair are aggregated to form the confusion matrix. Notably, to eliminate the influence of data imbalance, we ensure that each category is represented with an equal number of samples.

\subsection{ECG Inherent Characteristics}

% \begin{wrapfigure}[15]{r}{0.55\textwidth}
% \vspace{-1.em}
% \centering
% \includegraphics[width=0.5\textwidth]{figs/exp1.pdf}
% \caption{\small Hierarchical Composition of ECG Signals Highlighting Dominant and Subtle Features.}
% \label{fig:exp1}
% \vspace{0.1em}
% \end{wrapfigure}

\begin{figure}[h!]
% \vskip 0.2in
\begin{center}
\centerline{\includegraphics[width=\columnwidth]{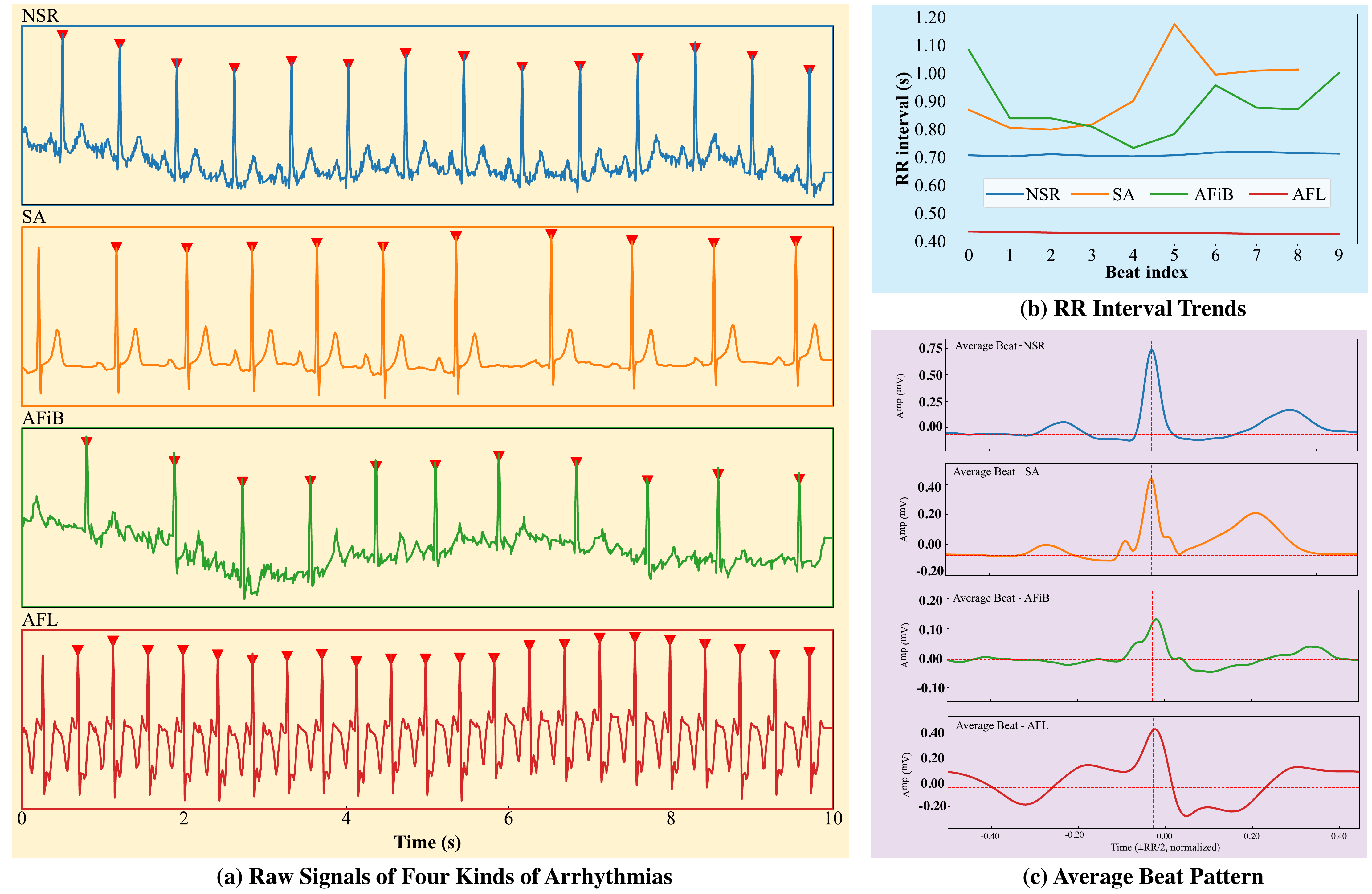}}
\caption{\small{ECG morphological patterns and rhythm characteristics across four cardiac conditions. (a) the overall ECG signals of NSR, SA, AFiB, and AFL are presented, capturing their raw waveform pattern. (b) visualizes the distribution of RR intervals for each condition, highlighting the rhythmic differences among them. (c) displays the average waveform of a single heartbeat for each condition.}}
\label{fig:exp1}
\end{center}
\vskip -0.4in
\end{figure}

As illustrated in Fig. \ref{fig:explain}, ECG signals encapsulate both dominant coarse-grained patterns and subtle fine-grained features, all of which are critical considerations for clinicians. To provide a clear understanding of ECG composition and analyze the characteristics of its components from a medical perspective, we conduct a qualitative exploration to identify the features that are inherently more challenging for the model. This analyses aims to shed light on how simplicity bias (SB) manifests in ECG signal interpretation. As shown on the left of Fig. \ref{fig:exp1}, the ECG signals exhibit distinct morphological patterns across different cardiac conditions, revealing clear differences in dominant and subtle features. 

To further highlight the importance of both dominant and subtle characteristics, from the clinical perspective, we consider four representative conditions: Sinus Arrhythmias (SA), Atrial Fibrillation (AFiB), Normal Sinus Rhythm (NSR) and Atrial Flutter (AFL). As shown in the top-right part of Fig. \ref{fig:exp1}, the relationship between RR intervals and heartbeats is visualized for these four states. It is evident from the figure that dominant features enable clear differentiation between AFiB and NSR. However, when it comes to distinguishing AFiB from SA, relying solely on dominant features proves insufficient due to overlapping rhythm characteristics. As shown in the bottom-right of Figure \ref{fig:exp1}, the waveform of a single heartbeat for each of the four conditions is presented. It is observed that variations in subtle waves provide clear differentiation between SA and AFiB.

% \begin{figure}[h]
% \centering
% \begin{subfigure}[t]{0.55\columnwidth}
%     \includegraphics[width=\linewidth]{figs/exp1.pdf}
%     \caption{\small{ Detailed Illustration of ECG Components.}}
%     \label{fig:exp1}
% \end{subfigure}
% \hfill
% \begin{subfigure}[t]{0.44\columnwidth}
%     \includegraphics[width=0.9\linewidth]{figs/exp2.pdf}
%     \caption{\small{ Visualization of RR Intervals Across Different Heart Conditions.}}
%     \label{fig:exp2}
% \end{subfigure}
% \caption{\small{ECG Composition and Diagnostic Characteristics Across Heart Conditions.}}
% \label{fig:cam_cf}
% \vskip -0.12in
% \end{figure}

% \subsection{Additional Qualitative Results Validating the Simplicity Bias Phenomenon}
% In this section, we present additional visualization results to illustrate the impact of simplicity bias on different methods.

\section{Experimental Setup}

\subsection{Downstream Tasks and Evaluation Metrics }
\label{appendix:expsetup1}

To comprehensively assess the effectiveness and generalizability of our proposed ECG representation learning method, we evaluate it across three downstream tasks: \textbf{multi-label classification}, \textbf{QRS segmentation}, and \textbf{mortality forecasting}. These tasks are carefully selected to reflect core clinical applications of ECG interpretation and to test the model’s ability to extract meaningful representations across a range of physiological timescales and abstraction levels. Together, they probe the model’s sensitivity to local waveform structures, its capacity for fine-grained diagnostic discrimination, and its ability to encode long-range prognostic information.

In the\textbf{\textit{ multi-label classification task}}, the objective is to identify multiple co-occurring cardiac conditions from a single ECG recording. This task evaluates the model’s ability to recognize both rhythm abnormalities and morphological deviations, reflecting its fine-grained understanding of waveform patterns and inter-lead dependencies. 
We apply a thresholding strategy to the model’s output logits to obtain binary label predictions. Performance is quantified using Area Under the Receiver Operating Characteristic Curve (AUROC), F1 score, and Accuracy, which together capture discrimination ability, sensitivity to label imbalance, and overall prediction fidelity across diverse diagnostic classes.

In the\textbf{\textit{ QRS segmentation task}}, we evaluate the model’s precision in identifying R-peak positions, which are central to rhythm analyses, beat-to-beat variability, and interval measurement. Candidate R-peak locations are first extracted from the model’s activation output using a peak-prominence detection strategy. We then apply clinically informed constraints based on the refractory period and typical QRS width to eliminate physiologically implausible peaks. The final predicted R-peak is defined as the midpoint between consecutive surviving candidates. A prediction is considered correct if it lies within a ±75 ms window of the annotated ground truth \cite{aami1998ec38}. Evaluation metrics include Sensitivity (Se), Positive Predictive Value (PPV), and F1 score, reflecting the model’s ability to detect true beats while minimizing false positives.

For the \textit{\textbf{mortality forecasting task}}, we formulate ECG-based prognosis as a survival analyses problem, using the full 10-second 12-lead ECG signal to predict patient mortality outcomes. This task emphasizes the model’s ability to capture long-term prognostic markers embedded in the raw waveform, such as repolarization abnormalities or autonomic tone, which may not be obvious in standard classification settings. We evaluate performance using the concordance index (C-index), which measures the consistency between predicted risk scores and actual event ordering, and the Brier score, which assesses the calibration of predicted survival probabilities. This setup allows us to quantify the model’s utility in risk stratification and its potential for integration into predictive clinical workflows.

\subsection{Downstream Datasets Description}
The summary of downstream task datasets is presented in Table \ref{tab:downstream-data}.

\begin{table}[t!]
\vspace{-0.8em}
\caption{Overview of ECG datasets used across downstream tasks. Attributes prefixed with \# denote quantity-based statistics.}
\label{tab:downstream-data}
\centering
\footnotesize
\renewcommand{\arraystretch}{1.1}
\setlength{\tabcolsep}{4.85pt}
\begin{tabular}{l|cccc|c|c}
\toprule
\multirow{2}{*}{\textbf{Method}}
& \multicolumn{4}{c|}{\textbf{Classification}} 
& \textbf{Segmentation} 
& \textbf{Forecasting} \\
\cline{2-7}
& \textit{PTB-XL} & \textit{Ningbo} & \textit{Chapman} & \textit{CPSC2018} 
& \textit{CPSC-QRS} 
& \textit{Sami-Trop} \\
\midrule
\# Recordings      & 21,837 & 45,152 & 10,646 & 9,364 & 9,364 & 1,959 \\
\# Patients        & 18,885 & 34,905 & 10,646 & 6,877 & 6,877 & 1,631 \\
Sampling Rate (Hz) & 500    & 500    & 500    & 500   & 500   & 400 \\
Duration (s)       & 10     & 10     & 10     & 6--60 & 10    & 7 or 10 \\
\# Labels          & 71     & -      & 67     & 9     & R-peak location  & Mortality \\
\bottomrule
\end{tabular}
\vskip -0.1in
\end{table}

\paragraph{Classification} \leavevmode
\begin{itemize}[leftmargin=2em]
    \item \textbf{{\textit{PTB-XL}}} \cite{strodthoff2020deep}A large-scale clinical dataset comprising 21,837 12-lead ECG recordings from 18,885 patients, each 10 seconds in duration and sampled at 500 Hz. Annotated by up to two cardiologists, the dataset includes 71 diagnostic labels conforming to the SCP-ECG standard, grouped into 5 superclasses and 44 sub-classes, making it the most popular resource for multi-label classification tasks. Available at: \href{https://physionet.org/content/mimiciv/2.2/}{PhysioNet (MIMIC-IV v2.2)}.

    \item \textbf{\textit{Chapman}} \cite{zheng202012} Includes 10,646 ECG recordings, each 10 seconds long and sampled at 500 Hz, collected from patients at Shaoxing People’s Hospital in China. It features annotations for 11 common rhythms and 67 additional cardiovascular conditions, providing a rich dataset for arrhythmia classification and rhythm analyses. Available at: \href{https://physionet.org/content/ecg-arrhythmia/1.0.0/}{PhysioNet (Chapman v1.0)}.

    \item \textbf{\textit{Ningbo}} \cite{zheng2020optimal} Comprising 45,152 12-lead ECG recordings from 34,905 patients, each 10 seconds in duration and sampled at 500 Hz, this dataset was collected from Ningbo First Hospital in China. It includes annotations for various cardiac abnormalities, offering a valuable resource for developing and evaluating arrhythmia detection algorithms. Available at: \href{https://physionet.org/content/ecg-arrhythmia/1.0.0/}{PhysioNet (Ningbo v1.0)}.

    \item \textbf{\textit{CPSC2018}} \cite{liu2018open} The China Physiological Signal Challenge 2018 dataset includes 6,877 12-lead ECG recordings, each ranging from 6 to 60 seconds and sampled at 500 Hz. The dataset encompasses nine arrhythmia classes, providing a benchmark for multi-class classification tasks. Available at: \href{https://physionet.org/content/challenge-2020/1.0.2/}{PhysioNet (CPSC 2018 v1.0.2)}.
       
\end{itemize}

\paragraph{Segmentation} \leavevmode
\begin{itemize}[leftmargin=2em]

    \item \textbf{\textit{CPSC2018-QRS}} \cite{gao2023ecg}  Derived from the CPSC2018 dataset, this subset is annotated with precise QRS complex locations, facilitating the evaluation of ECG segmentation algorithms, particularly for QRS detection tasks.
\end{itemize}

\paragraph{Forecasting} \leavevmode
\begin{itemize}[leftmargin=2em]
    \item \textbf{\textit{Sami-Trop}} \cite{ribeiro2021sami} An NIH-funded prospective cohort study comprising 1,959 12-lead ECG recordings from patients with chronic Chagas cardiomyopathy in Brazil. Each recording is sampled at 400 Hz and annotated with mortality outcomes and age information, enabling research into ECG-based survival prediction and disease progression modeling. Available at: \href{https://zenodo.org/records/4905618}{Zenodo (Sami-Trop v1.0.0)}.
\end{itemize}

% \subsection{Data Pre-Processing}
% All ECG recordings are first standardized to a consistent lead configuration, following the clinical 12-lead order: I, II, III, aVR, aVL, aVF, V1, V2, V3, V4, V5, V6. This ensures spatial alignment across datasets and preserves inter-lead dependencies critical for morphological and rhythm analyses. Signals are then resampled to 100 Hz to unify temporal resolution and reduce computational complexity. For recordings longer than 10 seconds, we extract non-overlapping consecutive 10-second segments; segments shorter than 10 seconds are discarded to maintain consistency in input length. A third-order Butterworth bandpass filter (0.5–45 Hz) is applied to each segment to remove baseline drift and high-frequency noise while preserving the clinically relevant frequency band. Segments containing NaN or Inf values are retained if such values constitute no more than 5\% of total samples, in which case they are replaced with zeros; otherwise, the segment is excluded. Finally, all valid segments are min-max normalized to the [0, 1] range on a per-segment basis, ensuring consistent amplitude scaling across patients and acquisition conditions.
\subsection{ECG Preprocessing Pipeline}
\label{appendix:preprocessing}

We adopt a unified preprocessing pipeline for both pre-training and downstream tasks to ensure consistency across datasets and experimental stages. The steps are as follows:

\begin{itemize}[leftmargin=2em]
    \item \textbf{Lead Standardization.} All ECG recordings are reordered to follow the canonical 12-lead configuration: I, II, III, aVR, aVL, aVF, V1–V6. This harmonizes spatial structure across datasets and preserves inter-lead dependencies essential for rhythm and morphological analyses.
    
    \item \textbf{Temporal Resampling.} Signals are resampled to 100 Hz to unify temporal resolution across sources and reduce computational overhead, while retaining clinically relevant waveform features.
    
    \item \textbf{Segment Extraction.} For recordings longer than 10 seconds, we extract non-overlapping consecutive 10-second segments. Recordings shorter than 10 seconds are discarded to maintain a consistent input length.

    \item \textbf{Missing/Invalid Values.} Segments containing NaN or Inf values are retained only if such values constitute less than 5\% of the total samples; in these cases, invalid values are replaced with zeros. Segments exceeding this threshold are discarded.
    
    \item \textbf{Filtering.} A third-order Butterworth bandpass filter (0.5–45 Hz) is applied to remove baseline wander and high-frequency noise, preserving the diagnostic frequency range relevant to clinical interpretation.
    
    \item \textbf{Normalization.} Each segment is min-max normalized to the [0, 1] range independently to ensure consistent amplitude scaling across patients and acquisition devices.
\end{itemize}

This pre-processing protocol is applied uniformly to all datasets used in both the self-supervised pre-training stage and the supervised downstream tasks, enabling standardized representation learning across heterogeneous ECG sources.

\subsection{Empirical Settings}
\label{esetting}
For a fair comparison, following \cite{zhang2022maefe, wang2023unsupervised, na2024guiding}, all methods are optimized using AdamW with $\beta_1 = 0.9$, $\beta_2 = 0.999$, a learning rate of 0.001, and a batch size of 1024. The total pretraining consists of 600 epochs with a warm-up period of 40 epochs. After pre-training, we fine-tune the pre-trained encoder on downstream datasets using the AdamW optimizer and a cosine learning rate schedule. Default hyperparameters include momentum ($\beta_1, \beta_2) = (0.9, 0.999)$, weight decay of 0.05, learning rate of 0.001, and a batch size of 512. We train for a maximum of 200 epochs with a warm-up period of 5 epochs. Notably, all methods are retrained under identical settings. Our TFF module, built upon the ViT-Base backbone, we incorporate 12 temporal and 12 frequency-domain blocks to enable time-frequency fusion. For MGPR, since the decoder in ViT is typically lighter than the encoder, we align encoder features with decoder blocks to maintain consistent granularity. Additionally, the settings of the compared methods are kept consistent with those of ST-MEM \cite{na2024guiding}. All experiments are conducted using two NVIDIA A100 GPUs.

\section{ECG-adapted Patch Embedding}
\label{appendix:patch}

A key design challenge in applying Transformer-based architectures to ECG signals lies in how to tokenize the input—i.e., how to divide continuous multi-lead waveforms into discrete patches for sequence modeling. Existing approaches often flatten or concatenate the 12-lead ECG signals along the temporal axis and treat the channels as independent features. While simple, this practice ignores the inherent spatial correlation among leads, which is clinically important for interpreting wave propagation and axis deviation, and also leads to an unfavorable trade-off between patch granularity and computational cost. Specifically, setting a small patch size increases the number of tokens, which exacerbates the quadratic complexity of the Transformer’s self-attention mechanism. Conversely, using large patches sacrifices fine-grained temporal resolution, impairing the model’s ability to detect subtle waveform variations critical for diagnosis.

To overcome this dilemma, we adopt a hybrid approach that leverages the strength of convolutional networks for local pattern extraction and cross-channel interaction, followed by Transformer-based global sequence modeling. Specifically, we employ a lightweight 1D ResNet \cite{wang2017time} with convolutional kernel size 5 as a patch embedding module, which processes the input ECG signal of shape $B \times 12 \times 1000$ ($batch size \times  leads \times time$). The ResNet encoder reduces the signal to an output of shape $B \times 256 \times 125$, where $125$ represents the number of patches (i.e., temporal tokens) and $256$ denotes the embedding dimension for each patch. This architecture allows the model to preserve inter-lead dependencies and extract hierarchical local features, while producing a manageable number of informative patches suitable for Transformer-based attention modeling.

This convolutional tokenization strategy offers several advantages: (1) it preserves the anatomical structure of multi-lead ECGs; (2) it reduces the computational overhead compared to naive token splitting; and (3) it enables fine-to-coarse hierarchical representation learning that aligns with the physiological organization of cardiac signals. As a result,  our patchifying approach strikes an effective balance between representation granularity, model efficiency, and clinical relevance. The quantitative experimental results are provided in detail in Appendix \ref{appendix:ablation3}.

\section{Additional Experiments}

\subsection{Effectiveness of ECG-adapted Patch Embedding}
\label{appendix:ablation3}
In this section, we conduct experiments to evaluate the effectiveness of the ECG-adapted patch strategy. We compare the standard MAE patch embedding with our ECG-adapted version through downstream classification experiments on the PTB-XL \cite{strodthoff2020deep} and CPSC2018 \cite{liu2018open} datasets, as shown in Table \ref{tab:app_sf}. It is evident that the performance slightly drops without the improved patch strategy, underscoring the effectiveness of our ECG-tailored patch embedding. Nevertheless, even without this enhancement, our method still outperforms the state-of-the-arts. 
% Besides, segmentation results in Table \ref{tab:segmentation_forecasting} greatly highlights that methods use o

\begin{table}[h]
\vspace{-0.8em}
\caption{\small Ablation study of different patch strategies on classification tasks (all metrics in \%). The results are reported on PTB-XL and CPSC2018. Entries marked with \textbf{*} indicate the use of the original patch strategy in standard MAE.}
\label{tab:app_sf}
\centering
\footnotesize
\renewcommand{\arraystretch}{1.1}
\setlength{\tabcolsep}{5pt}
\begin{tabular}{c|ccc|ccc}
\toprule
\multirow{2}{*}{\textbf{Method}} & \multicolumn{3}{c|}{\textit{PTB-XL}} & \multicolumn{3}{c}{\textit{CPSC2018}} \\
 & Accuracy & AUROC & F1 & Accuracy & AUROC & F1 \\
\midrule
MAE\textsuperscript{*}           & 88.64 & 92.21 & 73.27 & 96.21 & 96.50 & 83.72 \\
MAE           & 88.81 & 93.00 & 73.36 & 96.60 & 97.23 & 84.50 \\
Ours\textsuperscript{*}           & 89.45 & 93.41 & 74.55 & 97.36 & 97.83 & 85.00 \\
Ours          & 89.60 & 93.60 & 74.70 & 97.55 & 98.03 & 85.19 \\
\bottomrule
\end{tabular}
\vskip -0.1in
\end{table}

\section{Related Works}
\subsection{ECG Analyses}
The analyses of subtle ECG features, such as P and T waves, has long been challenging due to their low amplitude and variability. Early approaches focused on time-domain morphology enhancement \cite{clifford2012signal, wang2023adversarial}, template matching \cite{homaeinezhad2014correlation}, and wavelet-based methods \cite{banerjee2012delineation, karimipour2014real}, but often struggled under noisy or abnormal conditions. Later methods, including manual annotation and fixed-window segmentation \cite{ye2012heartbeat, rajesh2017classification}, improved robustness but lacked scalability. With the rise of deep learning, ECG analyses has shifted toward end-to-end modeling using multi-scale filters \cite{gao2023ecg, wang2024optimizing}, attention mechanisms \cite{islam2023hardc, zhang2023self}, and tailored loss functions \cite{golany2021ecg}. These methods have advanced performance across tasks but still suffer from simplicity bias, over-focusing on dominant features like QRS complexes while neglecting subtle patterns such as P and F waves \cite{reyna2021will, reyna2022issues, na2024guiding, zhang2022maefe}. Addressing this requires developing more comprehensive representations sensitive to both large-scale and fine-grained ECG information.

\subsection{Simplicity Bias in Pattern Recognition}

The concept of simplicity bias has garnered increasing attention for its profound influence on model optimization and generalization. Despite the capacity of neural networks to memorize random labels, they often generalize well due to training dynamics that implicitly favor simpler solutions \cite{zhang2021understanding, arpit2017closer, shah2020pitfalls}. While this preference facilitates efficient learning, it limits the ability to capture complex patterns—particularly in imbalanced scenarios such as long-tailed distributions, where dominant classes are learned more easily while minority class intricacies are neglected \cite{wei2025delving}. In transformers, this bias manifests as an early reliance on simple token interactions, and is further reinforced by techniques like sharpness-aware minimization, which encourage smoother solutions at the cost of fine-grained structure \cite{rende2024a, gatmiry2024simplicity}. Empirical evidence, such as Hessian spectrum analyses \cite{singh2021analytic}, shows that fitting random labels requires higher model complexity, highlighting how simplicity bias shapes the learning trajectory \cite{zhang2021understanding, arpit2017closer}. Although simplicity bias can promote robustness and efficiency in standard tasks, it poses challenges for adversarial robustness, minority class modeling, and structured data understanding. Addressing these issues calls for methods such as re-weighting, adversarial training, and architecture-level adaptation to balance inductive bias toward simplicity with the capacity for complex pattern modeling \cite{tsipras2018robustness}.

\subsection{Self-Surpervised Learning on ECG}

Self-supervised learning (SSL) methods have gained traction in ECG analyses, offering a promising pathway to improve feature learning. Early SSL methods treated ECG signals as image-like data, transforming them into spectrograms or scalograms for input into vision-based models \cite{vaid2023foundational}. Recent advancements have focused on time-domain SSL, with tasks like masked reconstruction encouraging models to capture global rhythms and local waveforms by reconstructing missing signal segments \cite{zhang2022maefe, na2024guiding}. Contrastive learning has further improved robustness through ECG-specific augmentations such as lead masking and time-warping \cite{lai2023practical, wang2023adversarial, zhang2023self}. Additionally, ST-MEM integrate spatial dependencies between 12-lead signals with temporal dynamics, enabling the learning of inter-lead correlations and fine-grained temporal features \cite{na2024guiding}. These advancements have shown improved performance in tasks like arrhythmia classification. However, existing SSL methods often fail to address simplicity bias explicitly, leaving room for improvement in the learning of subtle but diagnostically essential ECG features. In this work, we propose a novel reconstructed SSL framework that explicitly targets SB, combining multi-grained temporal and frequency-domain features to enhance generalization and diagnostic accuracy.

\end{document}